\documentclass[fleqn,usenatbib]{mnras}

\usepackage{newtxtext,newtxmath}
\usepackage{multirow}
\usepackage{xcolor}

\usepackage[T1]{fontenc}

\DeclareRobustCommand{\VAN}[3]{#2}
\let\VANthebibliography\thebibliography
\def\thebibliography{\DeclareRobustCommand{\VAN}[3]{##3}\VANthebibliography}

\usepackage{graphicx}	
\usepackage{amsmath}	

\newcommand{\kmpspmpc}{\,kms$^{-1}$Mpc$^{-1}$}
\newcommand{\T}{$\text{T}_{\text{inf}}\,$}
\newcommand{\V}{$V_{\text{max}}$}

\title[Star Formation Quenching \& Morphological Transformation]{Satellite quenching and morphological transformation of galaxies in groups and clusters}

\author[M. Oxland \& L. C. Parker]{
M. Oxland,$^{1}$\thanks{E-mail: oxlandm@mcmaster.ca}
L. C. Parker,$^{1}$
R. R. de Carvalho,$^{2}$ 
V. M. Sampaio,$^{2,3}$
\\
$^{1}$Department of Physics and Astronomy, McMaster University, Hamilton ON L8S 4M1, Canada\\
$^{2}$NAT-Universidade Cidade de S\~ao Paulo, 01506-000, S\~ao Paulo, SP, Brazil\\
$^{3}$School of Physics and Astronomy, University of Nottingham, University Park, Nottingham NG7 2RD, UK\\
}

\date{Accepted XXX. Received YYY; in original form ZZZ}

\pubyear{2023}

\begin{document}
\label{firstpage}
\pagerange{\pageref{firstpage}--\pageref{lastpage}}
\maketitle

\begin{abstract}
We investigate the role that dense environments have on the quenching of star formation and the transformation of morphology for a sample of galaxies selected from the Sloan Digital Sky Survey. We make a distinction between galaxies falling into groups ($ 13 \leq \log{(M_{\text{halo}}/M_{\sun})} < 14$) and clusters ($\log{(M_{\text{halo}}/M_{\sun})} \geq 14$), and compare to a large sample of field galaxies. Using galaxy position in projected phase space as a proxy for time since infall, we study how galaxy specific star formation rate (sSFR) and morphology, parameterized by the bulge-to-total light ratio (B/T), change over time. After controlling for stellar mass, we find clear trends of increasing quenched and elliptical fractions as functions of infall time for galaxies falling into both groups and clusters. The trends are strongest for low mass galaxies falling into clusters. By computing quenching and morphological transformation timescales, we find evidence that star formation quenching occurs faster than morphological transformation in both environments. Comparing field galaxies to recently infalling galaxies, we determine there is pre-processing of both star formation and morphology, with pre-processing affecting star formation rates more strongly. Our analysis favours quenching mechanisms that act quickly to suppress star formation, while other mechanisms that act on longer timescales transform morphology through bulge growth and disc fading.
\end{abstract} 




\section{Introduction}
Galaxies have long been observed to be largely bimodal in their properties, where the majority of galaxies are either (1) blue in colour, gas-rich, and have high star formation rates (SFRs) or are (2) red in colour, gas-poor, and have low SFRs \citep[e.g.][]{Strateva2001, Kauffmann2003b, Kauffmann2003, Baldry2004, Balogh2004a, Balogh2004b}. It is generally accepted that these two distinct populations represent different stages in galaxy evolution, where blue spiral galaxies have younger stellar populations compared to red elliptical galaxies \citep{Gonzalez2015}. Galaxies which have properties intermediate to these two distinct classes populate a region known as the green valley, which typically contains galaxies undergoing star formation quenching and experiencing a transformation in morphology from spiral to elliptical \citep{Martin2007}. The green valley contains relatively few galaxies, suggesting that the quenching of star formation likely occurs quickly especially for elliptical galaxies \citep{Salim2014, Schawinski2014}.

Understanding the mechanisms driving star formation quenching and morphological transformation requires a large sample of galaxies which span a variety of stellar masses and environments. The quiescent fraction of galaxies is known to increase with stellar mass, where low mass galaxies are typically star-forming while higher mass galaxies are more quiescent \citep[e.g.][]{Peng2010, Wetzel2012}. The quiescent fraction also shows a clear trend with local environment and scales with halo mass, where more quenched galaxies are found in more massive haloes \citep[e.g.][]{Wetzel2012, Jeon2022}. Even within individual galaxy clusters, satellites closer to the cluster centre have lower star formation rates than those further out \citep{Balogh1998, Roberts2019}. These trends suggest there are both internal and environmental quenching mechanisms that act in transforming galaxies over cosmic time. 

When a galaxy is in isolation, galaxy evolution is driven by internal processes. Outflows from Active Galactic Nuclei (AGN) and supernovae feedback can eject cold gas out of the disc and/or heat central gas. This reduces the abundance of cold gas and subsequently suppresses star formation \citep[e.g.][]{Bower2006, Croton2006, Agertz2013}. The presence of galactic bars can drive gas into central regions, increasing central star formation \citep[e.g.][]{Hawarden1986, Wang2012, Cheung2013} and inducing bulge growth \citep{Kormendy2004}. The growth of a stellar bulge can further stabilize the disc against fragmentation, reducing star formation through morphological quenching \citep{Martig2009}. However, most galaxies in the universe do not live in isolation. 

The most common environment for galaxies in the local universe is small groups \citep[e.g.][]{Eke2005}. Galaxies living in these small groups and larger galaxy clusters are subject to external transformation mechanisms in addition to the internal ones listed above. For example, a satellite galaxy passing through the dense intracluster medium (ICM) may experience a wind that is strong enough to strip cold gas from the galactic disc via ram pressure stripping (RPS) \citep[e.g.][]{Gunn1972}. In some cases, RPS can first compress cold gas and trigger an episode of enhanced star formation prior to complete gas removal \citep{Vulcani2018, Roberts2020}. If ram pressure is strong enough, it may eventually strip away all the cold star-forming gas causing the galaxy to quench on relatively short timescales (< 1 Gyr; \citet{Quilis2000}). Whether RPS enhances or quenches star formation depends on the halo mass of disc galaxies and their inclination with respect to their direction of motion \citep{Bekki2014}. Other environmental effects may strip only the hot halo, or starvation may prevent the hot gas from cooling, leading to the quenching of star formation as the cold gas reservoir is no longer being replenished \citep{Larson1980,Peng2015, Trussler2020}. Dynamical interactions such as repeated nearby encounters between nearby galaxies \citep{Moore1996} or mergers can also affect satellite SFRs and morphologies. 

To further complicate the picture of galaxy evolution, the growth of structure in the universe is a hierarchical process where massive galaxy clusters are built up by the accretion of smaller galaxy groups \citep{McGee2009}. As a result, many satellite galaxies experience pre-processing in which they have previously been affected by a group environment prior to entering their current cluster \citep[e.g.][]{Fujita2004, delucia2012, Wetzel2015, Bianconi2018}. Observationally, \citet{Hou2014} found that the observed quiescent fraction in massive clusters is dominated by pre-processed subhaloes, a result in agreement with simulations \citep[e.g.][]{Bahe2013, McGee2009}. Therefore, constraining both pre-processing and environmental quenching mechanisms is imperative to elucidate the processes driving galaxy evolution in dense environments.

One useful parameter to study how galaxy properties change when evolving through dense environments is the infall time: the time since a galaxy first crossed the virial radius of its present-day host halo \citep{Pasquali2019}. The direct way to measure infall time is to use phase space, a 6D position and velocity space commonly used to study dynamical systems. Galaxies falling into groups/clusters have well defined trajectories through phase space (see fig. 1 of \citet{Rhee2017}), but observationally we are limited to projected phase space (PPS). PPS is constructed with projected quantities (i.e line of sight velocities and projected distances from the cluster centre), where infall times can be extracted from cosmological simulations. Different galaxy populations are known to reside in distinct regions of PPS \citep[e.g.][]{Gill2005, Oman2013, Rhee2017, Pasquali2019}, where recent infalling satellites have larger velocity offsets with respect to the cluster centre, while ancient infallers are typically virialized and have small line-of-sight velocities. Using location in PPS for a large sample of galaxies, we can study how galaxy properties change over time.  

In recent observational studies, galaxy properties have been shown to correlate strongly with position in PPS. \citet{Mahajan2011} found that at a given projected distance from the cluster centre, the stellar mass and star formation properties of galaxies depend on their absolute line of sight velocity with respect to the cluster centre. \citet{Muzzin2014} studied a population of poststarburst galaxies and showed that they have a distinct distribution in PPS when compared to quiescent and star-forming cluster galaxies. \citet{Barsanti2018} found that passive galaxies dominate the virilaized regions of PPS in both group and cluster environments. Analyses of PPS have been carried out using simulations as well, for example \citet{Jaffe2015} used cosmological simulations to study the effect of ram pressure and determined that there is a specific location in PPS where stripped galaxies are likely to be found. These examples showcase that PPS is a valuable tool when studying galaxy evolution. 

In this work, we use the relationship between location in PPS and infall time to provide direct measurements of how SFR and morphology change with time. We build on the previous work of \citet{Sampaio2022} by including an analysis for both group and cluster environments, and we measure the timescales associated with changes in SFR and morphology. We also compare our results to field galaxies enabling us to measure the amount of pre-processing affecting galaxies of different stellar masses and in different environments. Our paper is structured as follows: in Section \ref{data} we define the catalogues from which we extract our galaxy sample, in Section \ref{sec:methods} we explain the methods we use in setting up PPS and assigning infall times to our galaxies. In Section \ref{results} we present our results on how galaxy SFR and morphology change as a function of infall time for galaxies in groups and clusters. We discuss and interpret our results in Section \ref{discussion}, and present a summary of our results in Section \ref{summary/conclusions}. This paper assumes flat, $\Lambda$CDM cosmology with $\Omega_M = 0.3$, $\Omega_\Lambda = 0.7$, and $H_0 = 70$\kmpspmpc. 


\section{Data}
\label{data}
We select galaxies from the Sloan Digital Sky Survey Data Release 7 \citep[SDSS DR7;][]{Abazajian2009}, and limit our sample to local galaxies with z $\leq$ 0.1. 

\subsection{Galaxy groups/clusters from the Yang Group Catalogue}
\label{sec:yang}
In order to classify galaxies according to their environment, we adopt the Yang Group Catalogue \citep{Yang2007}. This catalog uses a modified version of the halo-based group finder algorithm to identify galaxy groups from the New York University Value-Added Galaxy Catalogue (NYU-VAGC; \citet{Blanton2005}). \citet{Yang2007} first identify the centres of potential groups and estimate their characteristic luminosity. The authors then estimate the mass, size, and velocity dispersion of the dark matter halo associated with each group, which they use to determine membership in redshift space. This process is then preformed iteratively until there are no further changes in group membership. Each group is assigned two halo mass estimates: one based on characteristic luminosity and the other based on characteristic stellar mass. We use the halo mass estimated from luminosity in this work, and define galaxy groups to have dark matter haloes with $13 \leq \text{log}(M_{\text{halo}}/M_{\sun}) < 14$, while galaxy clusters are those with $\text{log}(M_{\text{halo}}/M_{\sun}) \geq 14$. 

We select galaxy groups/clusters with at least 3 members (N $\geq$ 3) for our main sample. We also remove central galaxies from our analysis, which are those flagged as the most massive in each group/cluster in the Yang Group Catalogue. Unlike the traditional friends-of-friends method, this halo-based group finder can identify groups with only one member (N=1). This enables us to create a supplemental dataset of field galaxies which provides us with a control sample that are not currently experiencing environmental effects.

\subsection{SFRs and stellar masses from GALEX-WISE}
\label{sec:sfr}
We use SFR and stellar mass measurements from the medium-deep version 2 of the GALEX-SDSS-WISE Legacy Catalogue (GSWLC-2; \citet{Salim2016, Salim2018}), which is a value-added catalogue for SDSS galaxies within the GALEX footprint \citep{Martin2005}. GSWLC-2 derives stellar masses and SFRs from UV+optical+mid-IR SED fitting done using the CIGALE code \citep{Boquien2019}. For this work, we only keep galaxies with stellar masses of $\text{log}(M_{\star}/M_{\sun}) \geq 9.5$. We define quenched galaxies as those with $\log(\text{sSFR}) < -11 \text{yr}^{-1}$, a common value used in the literature, to distinguish star-forming and passive galaxies \citep{Salim2016}.

\subsection{Morphological classification}
\label{sec:meert}
To separate early-type from late-type galaxies, we use the bulge-to-total light ratio (B/T) as a measure of morphology. Unlike visual morphology, this measure is quantitative in nature as it is determined through surface brightness profile fitting. B/T is defined as the fraction of total flux coming from the bulge of the galaxy, so B/T = 0 is a pure disc galaxy while B/T = 1 is a completely bulge-dominated galaxy. 

We adopt the B/Ts presented in the UPenn SDSS PhotDec Catalog \citep{Meert2015}. Briefly, this catalogue performed 2D decompositions in the g, r, and i bands for each of the de Vaucouleurs, Sersic, de Vaucouleurs+Exponential and Seric+Exponetial models for ~$7\times 10^5$ galaxies in the SDSS DR7. We adopt the r band fits for the de Vaucouleurs bulge and Exponential disc fit for our analysis. The fits were done using the fitting routine GALFIT \citep{Peng2002} and analysis pipeline PYMORPH \citep{Vikram2010}, which generates a number of physically motivated flags. \citet{Meert2015} flag "bad galaxies" as those with catastrophically bad estimates of total magnitude and radius; we follow the author's recommendation and remove these galaxies from our sample. 

Previous studies found that a quantitative definition of early-type galaxies should be based on B/T and image smoothness (S) \citep[e.g.][]{Im2002, Tran2003}. We therefore apply a cut in S since elliptical galaxies are expected to be relatively smooth and symmetric. We use the r-band image smoothness parameters presented in \citet{Simard2011} as an additional parameter when defining elliptical galaxies. $S=R_T+R_A$, where $R_T$ and $R_A$ are calculated following Eq. 11 of \citet{Simard2002} and represent the amount of light in symmetric and asymmetric components, calculated from the fitting model. \citet{Simard2009} determined that a good definition of early-type galaxies that matches well to visual classifications are those with B/T $\geq$ 0.35 and S $\leq$ 0.075. However, in this work we are interested primarily in elliptical (E) galaxies but there is no clear way to distinguish E's from S0's using B/T alone. We use a slightly more conservative cut of B/T $\geq$ 0.5 with S $\leq$ 0.075 to define E's for this work. This choice removes some S0's, but we note our sample of ellipticals is still contaminated by S0's. 

We cross match the environment, SFR, and morphology catalogues together and remove any galaxies with invalid measurements of mass, SFRs, or halo mass (i.e. null/ NaN/ 0). In Section \ref{sec:methods} we explain further steps taken to define the set of galaxies in groups and clusters used in this analysis.  


\section{Methods}
\label{sec:methods}
\subsection{Projected Phase Space}
We locate galaxies in projected phase space (PPS) by plotting the line of slight velocity as a function of projected radial distance for each galaxy, both with respect to clustercentric coordinates. We normalize the x and y axes by the cluster's virial radius ($R_{180}$) and 1D velocity dispersion ($\sigma_{\text{1D}}$), respectively, in order to create a stacked version of PPS containing galaxies in groups and clusters. For consistency, we follow the \citet{Yang2007} definitions:
\begin{equation}
    \sigma_{\text{1D}} = 397.9 \text{km s}^{-1} \left(\frac{M_h}{10^{14}h^{-1}M_{\sun}} \right)^{0.3214}
    \label{vdis}
\end{equation}

\begin{equation}
    R_{180} = 1.26 h^{-1}\text{Mpc} \left(\frac{M_h}{10^{14}h^{-1}M_{\sun}}\right)^{1/3}(1+z_{\text{group}})^{-1}
    \label{virialr}
\end{equation}
Where $M_h$ is the halo mass and $z_{\text{group}}$ is the redshift of the group/cluster centre.

\begin{figure}
    \includegraphics[width=\columnwidth]{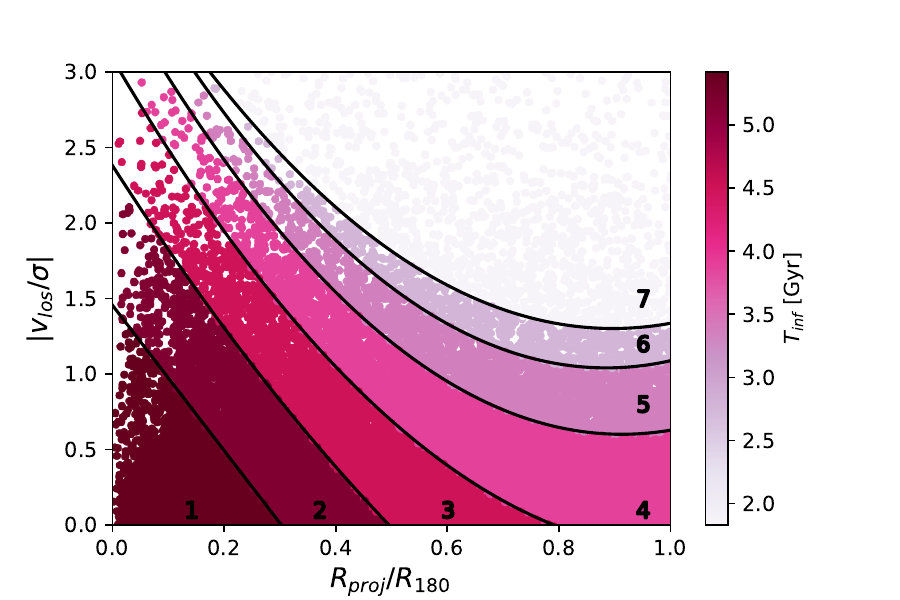}
    \caption{Projected phase space diagram for all our galaxies in groups and clusters. The colour bar distinguishes the 7 different regions determined by \citet{Pasquali2019}, where the corresponding infall times are provided in Table \ref{tab:Tinf}.}
    \label{fig:PPS}
\end{figure}

We adopt the "new zones" determined by \citet{Pasquali2019} to correlate position in PPS with a time since infall (\T). The authors define \T as "the time since a galaxy crossed for the first time the virial radius of the main progenitor of its present-day host environment". They used the Yonsei Zoom in Cluster Simulations (YZiCS; see \citet{Choi2017} for simulation details) to derive zones of constant mean infall time for galaxies in each region of PPS. The zones defined by \citet{Pasquali2019} are valid for $|v_{\text{los}}|/\sigma_{\text{1D}} \leq 3$ and $R_{\text{proj}}/R_{180} \leq 1$, which reduces our parent sample of galaxies in groups and clusters. The authors also divide PPS into 8 different zones, but we note their zone 7 is quite narrow so we combine their zones 7 and 8 into a single region with a corresponding mean infall time of 1.83 Gyr. For each zone, the number of galaxies (N), the mean infall time, and standard deviations are listed in Table \ref{tab:Tinf}. The PPS for our galaxy sample is shown in Fig. \ref{fig:PPS}.

\subsection{Populating the outskirts of projected phase space}
To improve the statistics within zones 6 and 7, we add additional galaxies to the outskirts of our groups and clusters that may not have been identified as members by the halo-based group finder of \citet{Yang2007}. We do so by finding all isolated galaxies (N=1) that are within 1 $R_{180}$ of a galaxy group centre, and then determine the line-of-sight velocity relative to this new cluster centre. We normalize these measurements by equations (\ref{vdis}) \& (\ref{virialr}), and keep those within $|v_{\text{los}}|/\sigma_{\text{1D}} \leq 3$ and $R_{\text{proj}}/R_{180} \leq 1$. This added 588 galaxies to our sample, 96\% of which are located within zones 6 and 7.

\begin{table}
    \centering
    \caption{Projected phase space zone number (see Fig. \ref{fig:PPS}), the number (N) of galaxies in each zone in my sample, the corresponding time since infall, and the associated standard deviations of \T. The numerical values are determined from simulations \citep{Pasquali2019}.}
	\label{tab:Tinf}
	\begin{tabular}{cccc}
		\hline
		Zone & N & $\text{T}_{\text{inf}}$ [Gyr] & $\sigma$(\T) [Gyr]\\
		\hline
		1 & 2296 & 5.42 & 2.51\\
		2 & 3763 & 5.18 & 2.60\\
		3 & 4602 & 4.50 & 2.57\\
		4 & 5632 & 3.89 & 2.34\\
		5 & 2136 & 3.36 & 2.36\\
		6 & 768 & 2.77 & 2.29\\
		7 & 1172 & 1.83 & 2.47\\
		\hline
	\end{tabular}
\end{table}

\subsection{\texorpdfstring{$V_{\text{max}}$}{TEXT} Volume Correction}
To correct for the Malmquist bias and the fact we are missing low-mass quenched galaxies since they are often below the SDSS detection limit, we apply a volume correction. Taken from the \citet{Simard2011} catalogue, \V is defined as the maximum volume over which the galaxy could be observed in $\text{Mpc}^3$.

Our final sample of galaxies in PPS contains 20369 galaxies, of which 12062 are in groups and 8307 are in clusters. For our field sample, we remove any galaxies in the Yang N=1 data set within $3R_{180}$ of a group or cluster. This results in a sample of 73474 field galaxies.

\section{Results}
\label{results}
It is well known that galaxies accreting into groups and clusters both decrease their star formation rate and experience a morphological transformation from disc to bulge-dominated. However, there is currently no clear consensus as to the physical mechanisms driving these changes, or the timescales associated with such processes. In this section, we explore both the quenched and elliptical fraction of galaxies as functions of time since infall to better constrain the physical mechanisms driving galaxy evolution through dense environments. Since galaxy properties are known to correlate with stellar mass and host halo mass \citep[e.g.][]{Kauffmann2003, Peng2010, Wetzel2012}, we separate our analysis into three stellar mass bins and two environments.

\subsection{Quenched Fraction}
\label{QF}
\begin{figure*}
    \includegraphics[width=\textwidth]{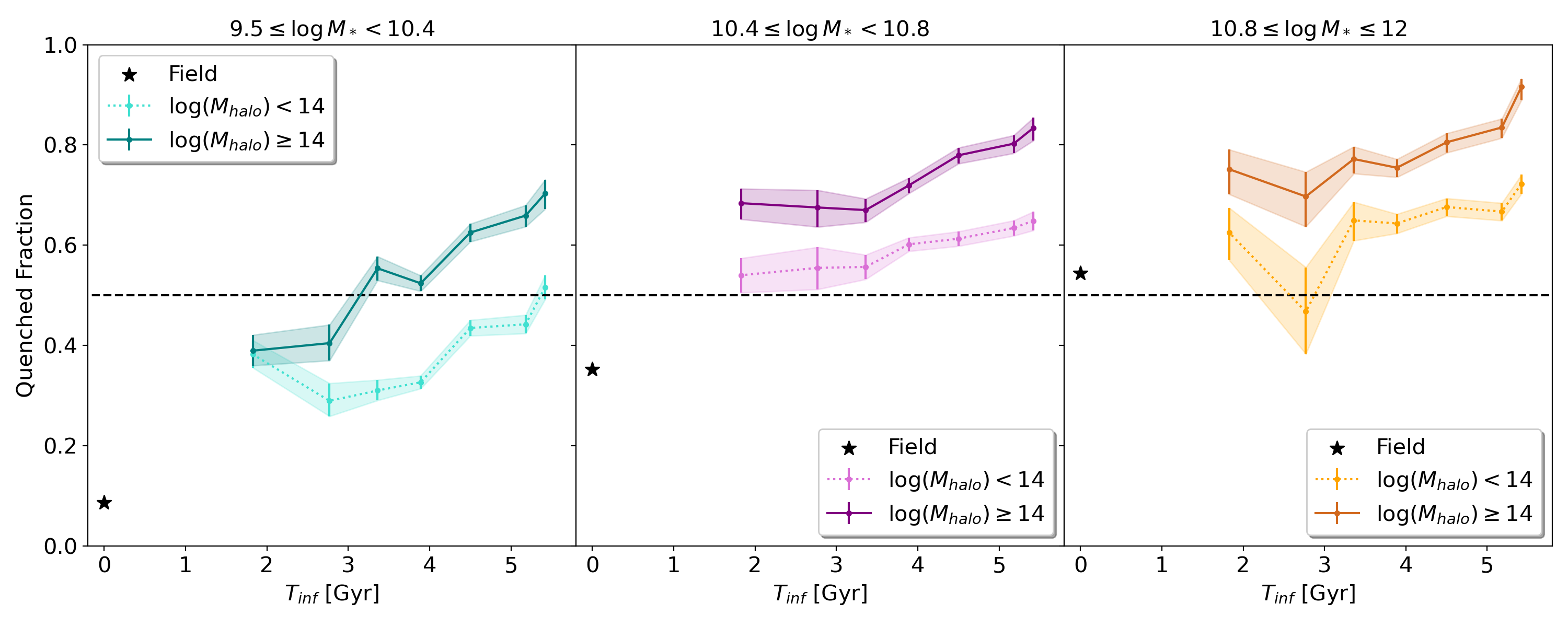}
    \caption{The quenched fraction of galaxies as a function of time since infall, where the quenched fraction is weighted by 1/\V as defined in Equation \ref{eqn: QF_eq}. The left most panel corresponds to low mass galaxies, while the middle and right most panels contain intermediate and high mass galaxies, respectively. The dotted lines in each panel represent the galaxies in groups, while solid lines are galaxies in clusters. The error bars corresponding to the 68$\%$ confidence intervals estimated from the beta distribution \citep{Cameron2011}. Black stars represent field galaxies, are artificially placed at \T = 0, and have uncertainties that are smaller than the marker size. Finally, the black dashed line represents the point at which 50$\%$ of galaxies in a given class are quenched. The uncertainty on the sSFR is not taken into account when calculating the QF.}
    \label{fig:QF}
\end{figure*}
In Fig. \ref{fig:QF} we show the quenched fraction as a function of time since infall, where \T is determined based on galaxy position in PPS (see Fig. \ref{fig:PPS}). Here, the quenched fraction (QF) is
\begin{equation}
    \text{QF} = \frac{\sum^{N_\text{q}}_{i=1}1/V_{\text{max}}}{\sum^{N_\text{total}}_{i=1}1/V_{\text{max}}}
    \label{eqn: QF_eq}
\end{equation}
where $N_\text{q}$ is the number of quenched galaxies ($\log(\text{sSFR}) < -11 \text{yr}^{-1}$) and $N_{\text{total}}$ is the total number of galxies in each \T bin \citep{Dickey2021}.

The three panels correspond to low ($9.5 \leq\text{log}(M_{\star}/M_{\sun}) < 10.4$), intermediate ($10.4 \leq\text{log}(M_{\star}/M_{\sun}) < 10.8$), and high ($10.8 \leq\text{log}(M_{\star}/M_{\sun}) < 12$) mass galaxies, chosen to have approximately the same number of galaxies in each mass bin. Within each panel we separate galaxies into those within groups ($13 \leq \text{log}(M_{\text{halo}}/M_{\sun}) < 14$) and clusters ($\text{log}(M_{\text{halo}}/M_{\sun}) \geq 14$). In addition, we artificially plot the quenched fraction of field galaxies at \T=0 to compare galaxies evolving through dense environments to those in the field. Note there are a minimum of 31 galaxies in each \T bin. The mean stellar masses are comparable within each \T bin for low, intermediate, and high mass galaxies.

The QF increases over time for galaxies falling into group and cluster environments, although the trends are stronger in clusters compared to groups. The population with the steepest slope are low mass galaxies in clusters, suggesting these galaxies are the most strongly affected by their environment. The dashed horizontal line at QF=0.5 indicates the separation between where most galaxies are star-forming (QF<0.5) or quiescent (QF>0.5), and is used to calculate the quenching timescales (see Section \ref{EFQF}). Interestingly, most low mass galaxies in groups and clusters are actively star-forming $\sim$1.8 Gyr after infall (left most point corresponding to PPS zone 7), while intermediate and high mass galaxies are typically already quenched (see discussion of pre-processing in Section \ref{sec:prepro}). We also note that the QF for intermediate and high mass galaxies appears to remain relatively flat until $\sim$3 Gyr, and then starts to increase. This could suggest these galaxies experience a delay phase followed by a decrease in their star formation rate, consistent with the delayed then rapid quenching scenario \citep[e.g.][]{Wetzel2013, Rhee2020}.

Unsurprisingly and in agreement with many previous works \citep[e.g.][]{Wetzel2012, Wetzel2013, Jeon2022}, the QF depends both on stellar mass and environment where the lowest fractions are found in the field and increase with stellar mass and halo mass. For completeness, we show the trend of sSFR as a function of \T in Fig. \ref{fig:sSFR}.

\subsection{Elliptical Fraction}
\label{EF}
\begin{figure*}
    \includegraphics[width=\textwidth]{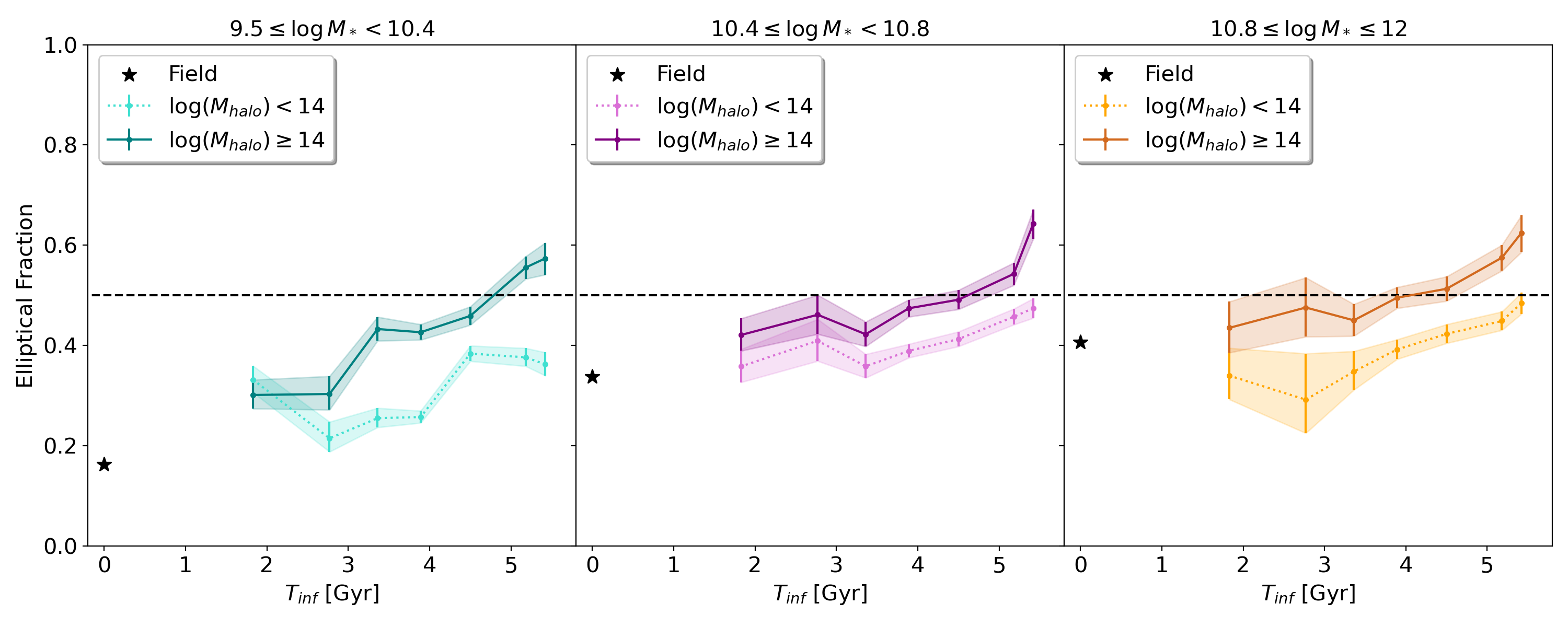}
    \caption{The elliptical fraction of galaxies as a function of time since infall, where the elliptical fraction is weighted by 1/\V. The separate panels and colours are the same as those defined in Fig. \ref{fig:QF}, with error bars corresponding to the 68$\%$ confidence intervals estimated from the beta distribution \citep{Cameron2011}. The error bars on the field populations are smaller than the black star marker, and the black dashed line represents the point at which 50$\%$ of galaxies in a given class are considered elliptical. The uncertainty on the B/T is not taken into account when calculating the EF.}
    \label{fig:EF}
\end{figure*}
 
In Fig. \ref{fig:EF} we plot the elliptical fraction as a function of \T. Here, the elliptical fraction (EF) is weighted by 1/\V and is calculated similarily to the QF, although $N_\text{q}$ in Equation \ref{eqn: QF_eq} is replaced by the number of elliptical galaxies ($N_\text{E}$; B/T $\geq$ 0.5 and S $\leq$ 0.075) within each \T bin.

The EF bears a striking resemblance to the QF in Fig. \ref{fig:QF}. For all galaxy populations, the EF increases as a function of \T and it depends on both stellar mass and halo mass. The steepest slope corresponds once again to low mass galaxies falling into clusters. The dashed horizontal line at EF=0.5 is the division between where most galaxies are disc dominated (EF<0.5) or elliptical (EF>0.5), and is used when calculating the morphological transformation timescales in Section \ref{EFQF}. In general, most galaxies fall into environments as disc-dominated systems and slowly become elliptical over time. However, galaxies in groups are typically disc dominated for the entire time they are a part of their group. The field is dominated by galaxies with disc morphologies.

We again note intermediate and high mass galaxies have EFs that remain quite flat up until $\sim$3 Gyr before increasing, supporting the delayed then rapid quenching scenario. For completeness, we provide a plot of the median B/T as a function of \T in Fig. \ref{fig:BT}. The overall trends are the same as the ones presented in this section. 

\subsection{What happens first: quenching or morphological transformation?}
\label{EFQF}
\begin{figure*}
	\includegraphics[width=\textwidth]{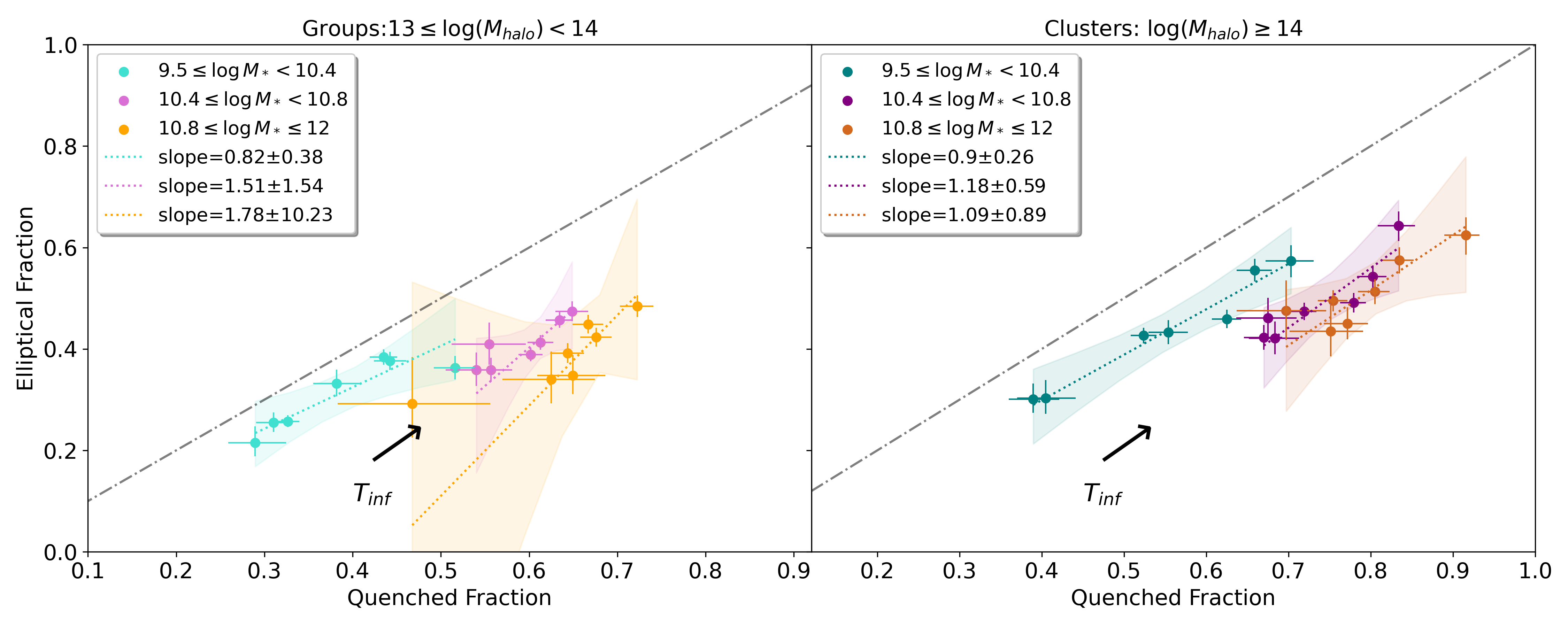}
    \caption{The elliptical fraction as a function of the quenched fraction, where both fractions are \V weighted. The left and right panels correspond to galaxies in groups and clusters, respectively, while galaxies in different stellar mass bins are represented by the different coloured points. Each data point corresponds to a specific time since infall, where \T increases from the bottom left to the top right as shown by the illustrative black arrow. The linear least squares fit to each population is shown by dotted lines, where the slope of each line is provided in the legend. The error bars on each data point correspond to the 68$\%$ confidence intervals estimated from the beta distribution \citep{Cameron2011}, and the shaded regions show the 90$\%$ confidence interval of the best-fitting line from LINMIX \citep{Kelly2007}. The dot-dash line shows the one-to-one line.}
    \label{fig:EFQF}
\end{figure*}
In the previous sections, we explored how the QF and EF change over time as galaxies fall into groups and clusters. In this subsection, we combine the results from Sections \ref{QF} and \ref{EF} in an attempt to address the question whether it is star formation quenching or morphological transformation that occurs first. We first do so by plotting the EF as a function of the QF, as shown in Fig. \ref{fig:EFQF}. Each data point corresponds to one specific \T, where \T increases from the bottom left to the top right as shown by the illustrative black arrow.
 
In Fig. \ref{fig:EFQF} we see that as galaxies fall into groups and clusters, the elliptical and quenched fractions both increase over time as expected. Within groups and clusters, there is significant overlap between intermediate and high mass galaxies. For each population we use the hierarchical Bayesian model LINMIX \citep{Kelly2007} to determine the best-fitting straight line (and it's associated $90\%$ confidence interval), taking into account the errors in both fractions. The slopes of the resultant best-fitting lines are provided in the legend. With large error bars in the best-fitting lines, it is difficult to determine from this analysis alone if star formation or morphology changes more quickly.

\begin{figure}
    \centering
    \includegraphics[width=0.45\textwidth]{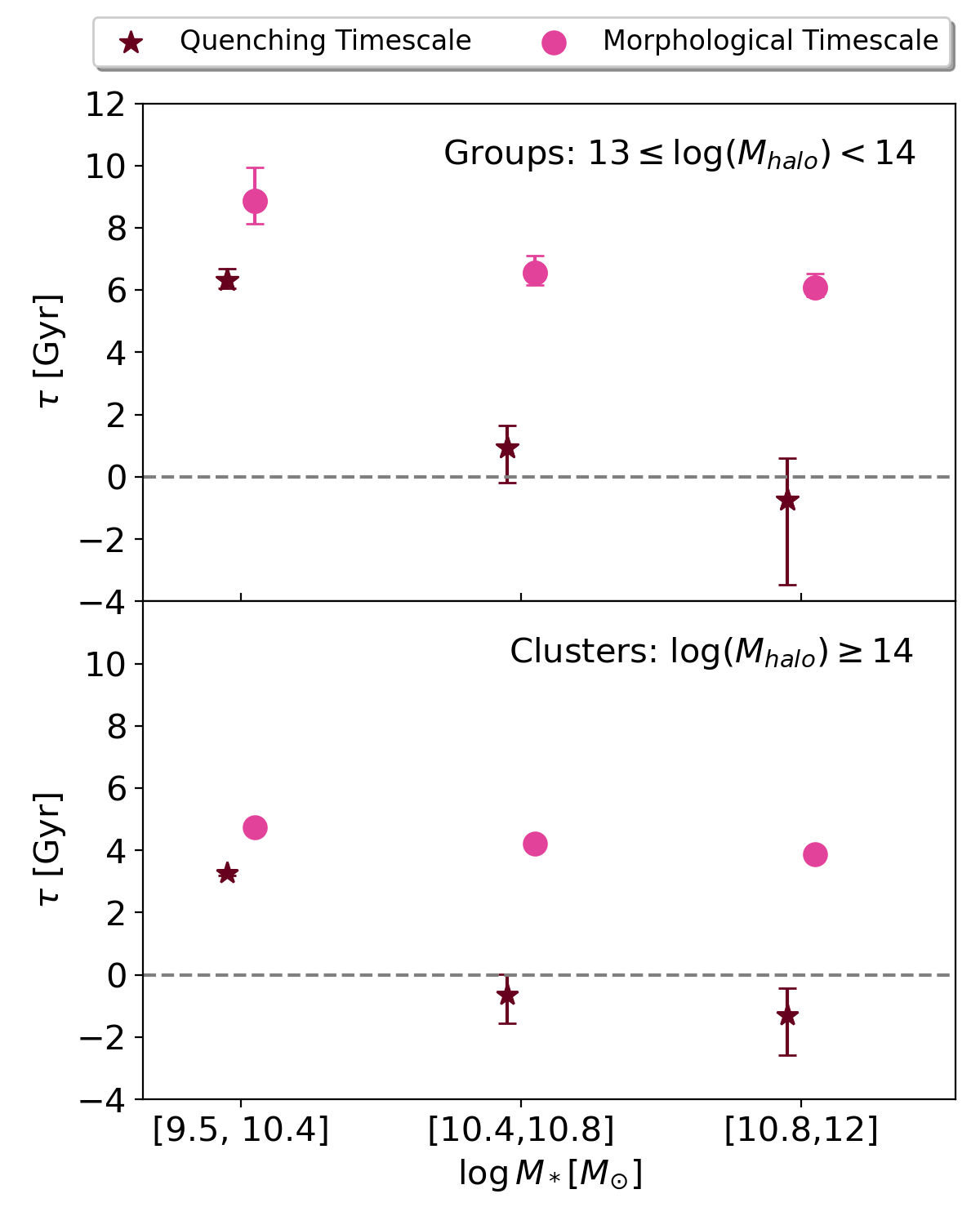}
    \caption{The quenching and morphological transformation timescales for low ($9.5 \leq\text{log}(M_{\star}/M_{\sun}) < 10.4$), intermediate ($10.4 \leq\text{log}(M_{\star}/M_{\sun}) < 10.8$), and high ($10.8 \leq\text{log}(M_{\star}/M_{\sun}) < 12$) mass galaxies in groups (top) and clusters (bottom). The quenching timescales are plotted as burgundy stars, while the morphological transformation timescales are plotted as pink circles. The errors are calculated from the formal uncertainty of the slope and y-intercept of the best-fitting line used to calculate the timescales (see text for details).}
    \label{fig:timescales}
\end{figure}

An alternative approach we take in addressing this question is to compute the physical timescales associated with quenching and morphological transformation. We define the quenching timescale as the time it takes for a specific galaxy population to have a QF=0.5. We do so by fitting a simple linear least squares fit to each sub-population in Fig. \ref{fig:QF}, taking into account the uncertainty on the QF, and determine at what \T value the quenched fraction crosses the dashed line. Similarly, we define the morphological transformation timescale as the \T at which a specific galaxy population has an EF=0.5, calculated from Fig. \ref{fig:EF}. Both timescales are plotted in Fig. \ref{fig:timescales}.

Within each stellar mass range we find that morphological transformation timescales are longer than quenching timescales, suggesting quenching happens prior to morphological transformation. Both timescales also decrease with increasing stellar mass, and the timescales are shorter in clusters when compared to galaxies of comparable mass in groups. Low mass galaxies in groups have extremely long morphological transformation timescales ($\sim$ 9 Gyr), suggesting these galaxies do not show significant signs of morphological transformation. Some galaxy populations also have negative quenching timescales and although this may seem nonphysical, it points to significant pre-processing of star formation prior to becoming satellites of their current environment. We discuss the effects of pre-processing further in Section \ref{sec:prepro}.

The \T associated with each PPS zone has a large standard deviation (see Table 1), which were not considered when calculating the linear fits. Given the large uncertainty in infall times estimated from projected phase space, the important results from Figs. \ref{fig:QF} and \ref{fig:EF} show the overall trend with infall time rather than the exact quantitative fractions at a given time since infall. We also note the individual measurement errors on stellar mass, sSFR and B/T are not taken into account in Figs. \ref{fig:QF}, \ref{fig:EF}, \ref{fig:EFQF} so the errors shown are likely underestimated. This should be kept in mind when interpreting our results.


\section{Discussion}
\label{discussion}
In this paper we investigate the environmental effect of groups and clusters on galaxy star formation rates and morphologies to help constrain satellite quenching and morphological transformation. Similar to previous work, our findings suggest that clusters have the most influence on galaxy sSFR and morphology, with low mass galaxies being the most susceptible to environmental effects. In terms of both the QF and EF, galaxies in both groups and clusters of all stellar masses show a clear offset from the field. This indicates galaxies have likely experienced pre-processing prior to entering their current host environment. In the following sections we interpret our results and discuss the physical mechanisms that may be driving galaxy evolution through dense environments.

\subsection{Evidence for star formation quenching prior to morphological transformation}
\label{sec:prior}
Our goal in this work was to answer the question of whether it is star formation quenching or morphological transformation that happens first for galaxies falling into groups and clusters. By looking at the elliptical fraction as a function of the quenched fraction in Fig. \ref{fig:EFQF}, we determine that both fractions are increasing as a function of infall time, but it is difficult to disentangle which is changing first due to large uncertainty. We expand upon this analysis by calculating the quenching and morphological transformation timescales (see Fig. \ref{fig:timescales}), where we find evidence that star formation quenching is changing prior to morphological transformation.

Our results agree with several recent studies that use PPS. \citet{Kelkar2019} used observational data from the ESO Distant Cluster Survey \citep{White2005} and found a correlation between stellar age and position in PPS, irrespective of morphology. They argue this is evidence star formation is suppressed earlier than morphology. \citet{Martinez2023} reached the same conclusion for cluster galaxies; they separated galaxies into five dynamical classes based on their position in PPS and compared the fraction of ellipticals and passives to an interloper population (see Section \ref{sec:Martinez} for a quantitative comparison). Our results disagree with those by \citet{Martig2009} and \citet{Sampaio2022}, who used cosmological simulations and SDSS observations, respectively. \citet{Martig2009} argue morphological transformation and star formation quenching occur simultaneously, and attribute the changes to morphological quenching. \citet{Sampaio2022} use galaxy position in PPS and T-Type as a measure of morphology to study how sSFRs and morphologies change over time in galaxy clusters. By plotting T-Type versus sSFR, they find a greater change in morphology compared to sSFR over a fixed time interval, arguing that this is evidence morphology is changing more rapidly than sSFR. 

Our analysis is similar to that of \citet{Sampaio2022} with a number of differences. Firstly, we use B/T as a measure of morphology instead of T-Type. T-Type is a measurement that comes from Convolutional Neural Networks calibrated with Galaxy Zoo 2 data \citep{Dominguez2018, Willett2013}, while B/T is calculated from photometric decompositions. While both are common morphology indicators, they are not perfectly correlated and may be the origin of our differing results. Secondly, the comparison of T-Type and sSFR is difficult to interpret as the two measures have differing units (see their fig. 10). Our choice to explore the QF versus the EF removes units, making comparisons between the changes in star formation and morphology more straight-forward. To make a direct comparison to  fig. 10 of \citet{Sampaio2022}, we provide a plot of B/T as a function of sSFR in Fig. \ref{fig:BTsSFR}. It appears sSFR is changing more quickly than morphology, but again note the difficultly in comparing changes in B/T and sSFR.

The quenching timescales we calculate provide us with a quantitative way to determine how SFRs change both as a function of stellar mass and environment. However, making direct comparisons to work in the literature is difficult due to the many ways quenching timescales are defined (see \citet{Cortese2021} for a review). Our quenching timescales are defined as the time from \T=0 to the time the population has a QF=0.5 (see Section \ref{EFQF}). As a result,  our timescales therefore represent the overall average quenching time for a population of galaxies rather than the time it takes for a single galaxy to become quenched. We find that our quenching timescales decrease with increasing stellar mass, and they are longer in groups compared to clusters.

Several works have tried to constrain satellite quenching timescales and have come to similar conclusions. \citet{delucia2012} used a semi-analytic model applied to the merger trees of the Millennium Simulation \citep{Springel2005}, computing quenching timescales of ~5-7 Gyrs and determining there is significant pre-processing of cluster galaxies. \citet{Wetzel2013} used both SDSS observations and cosmological simulations to study the star formation histories of galaxies falling into clusters. They proposed the "delayed-then-rapid" quenching model, where satellites experience little to no change in their SFRs for 2-4 Gyrs, after which they experience rapid star formation quenching ($\leq$ 1 Gyr). These timescales were found to be shorter at higher stellar masses, and they found the observed increase in the quiescent fraction with host halo mass is due to group pre-processing. \citet{Oman2016} used N-body simulations and SDSS galaxies in PPS to estimate quenching timescales. They accounted for pre-processing effects and found that all galaxies quench on their first infall, within 1 Gyr of their first pericentric passage. They find higher mass satellites quench earlier, with little dependence on host cluster mass. 

These results are in agreement with \citet{Rhee2020} who recently combined the YZiCS and SDSS observations to study the quenching of satellite galaxies in clusters ($\text{M}_{\text{halo}}>5\times10^{13} \text{M}_{\odot}$) using PPS. Rather than using the mean \T in each region of PPS, the authors separate PPS into square pixels where each pixel has a corresponding density distribution of \T. They then apply a flexible quenching model to calculate the ex-situ quenching timescale, the delay time, and the cluster-quenching timescale based on the SFR and \T of a given galaxy. \citet{Rhee2020} find that galaxies with $\text{M}_{*}>10^{9.5} \text{M}_{\odot}$ generally follow delayed-then-rapid quenching, where quenching begins outside of the cluster via mass quenching and pre-processing. Quenching is maintained during the delay phase for $\sim 2$ Gyr, followed by a rapid decline in SFR (e-folding timescale of $\sim 1$ Gyr). They also find a decline of quenching timescales with increasing mass, consistent with our analysis.

Although there is no one definition of quenching timescale used in the literature, the overall trend of decreasing quenching timescales with increasing stellar mass is clear. We find shorter quenching timescales in clusters compared to group environments, but explore the possibility this is due to pre-processing in Section \ref{sec:prepro}. 

We acknowledge that the specific timescales calculated here are dependent on the definition of EF, QF, and the thresholds chosen. For example, if we chose a more conservative cut of B/T$\geq$0.7 for our elliptical galaxy definition, the morphological transformation timescales all increase. Alternatively, the quenching timescales increase if we use a lower $\log(\text{sSFR})$ threshold to define quenched galaxies. We perform a stability test in Appendix \ref{stability test} to explore how our timescales change when different thresholds are chosen, and note that even with different thresholds our conclusion that star formation changes more quickly than morphology remains.

Since the numerical timescales calculated do change depending on the chosen thresholds, it is the overall trends with stellar mass, halo mass, and the comparison between the morphological transformation and quenching timescales that are meaningful, rather than the exact numerical timescales calculated. Since morphological transformation timescales depend on the measure of morphology used, future work is needed to better understand these differences.

\subsection{Evolution of cluster galaxies: more evidence for quenching prior to morphological change}
\label{sec:Martinez}
\begin{figure*}
    \includegraphics[width=\textwidth]{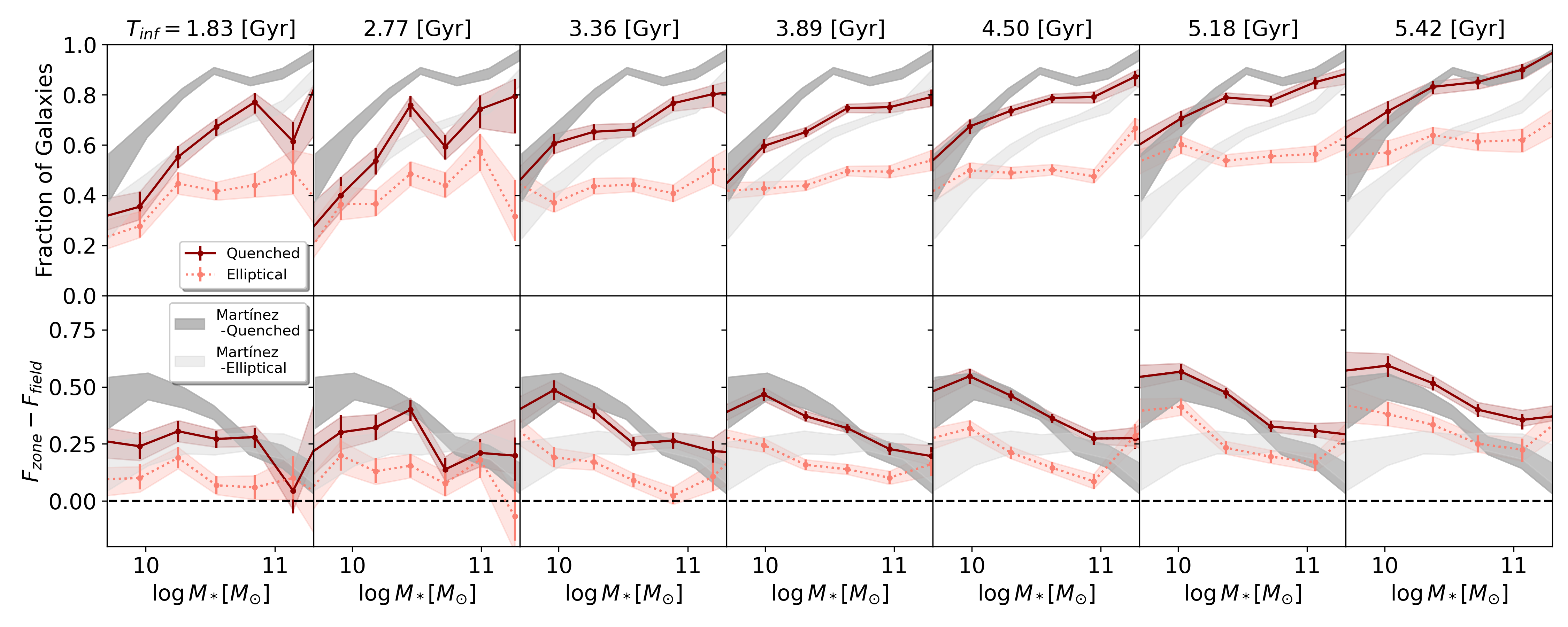}
    \caption{The top row shows the fraction of ellipticals (salmon coloured dotted line), and fraction of quenched galaxies (burgundy solid line) as a function of stellar mass, with each panel representing a specific \T labeled at the top of each plot. Each fraction is also \V weighted. Error bars correspond to the 68$\%$ confidence intervals estimated from the beta distribution \citep{Cameron2011}. In the bottom panel we show the difference of the elliptical fraction and quenched fraction relative to the fractions in our field sample. The horizontal black dashed line is plotted at $F_{\text{zone}}-F_{\text{field}}$=0. The elliptical fraction (dark grey) and passive fraction (light grey) for the cluster galaxies from fig. 7 of \citet{Martinez2023} are plotted in the background.}
    \label{fig:martinez}
\end{figure*}

Recently, \citet{Martinez2023} studied galaxies within bright X-ray clusters to explore how sSFR and morphology change over time. They apply the code ROGER \citep{delosRios2021}, which uses three different machine learning algorithms to classify galaxies into one of five orbital classes based on position in PPS. The five dynamical classes are cluster galaxies, recent infallers, backsplash galaxies, infallers, and interlopers. Based on the results of \citet{delosRios2021}, \citet{Martinez2023} adopt the K-Nearest Neighbour technique to compute class probabilities. Their sSFRs come from the MPA-JHU DR7 catalogue \citep{Brinchmann2004}, and their morphological classifications are from the Galaxy Zoo Project \citep{Lintott2008}. The authors use the probabilities of being elliptical ($P_E$) or spiral ($P_S$) from \citet{Lintott2011}, and consider a galaxy to be elliptical if $P_E > P_S$. 

\citet{Martinez2023} compare the fraction of ellipticals to the fraction of passives within each of their five predicted classes as a function of galaxy stellar mass (see their fig. 7). Interlopers are analogous to a field population, while the sequence of infallers, recent infallers, backsplash, and cluster galaxies corresponds roughly to time since infall. They find quenching is enhanced as soon as galaxies fall into clusters, while significant morphological transformations require longer periods of time. The authors come to this conclusion as it takes longer for the fraction of elliptical galaxies to significantly deviate from the field compared to the fraction of passives.

Since the progression from infallers to cluster galaxies traces galaxies through PPS, I can compare our work to the work of \citet{Martinez2023} using the 7 zones of our PPS (see Fig. \ref{fig:PPS}). We plot the elliptical fraction and quenched fraction of cluster galaxies as a function of stellar mass, where each panel corresponds to one \T bin in Fig. \ref{fig:martinez}. We plot the results for the cluster galaxy classification of \citet{Martinez2023} in the background, as this is the classification our PPS diagram overlaps the most with. We again consider a galaxy to have an elliptical morphology if B/T $\geq$ 0.5 and S $\leq$ 0.075, and to be quenched if $\log(\text{sSFR}) \leq -11 \text{yr}^{-1}$.

The top row of Fig. \ref{fig:martinez} shows both the EF and QF increase as a function of stellar mass. The fractions increase the longer a galaxy has been a part of a cluster, supporting the idea that environment has an effect on both SFR and morphology. Quenched fractions are also systematically larger than elliptical fractions over all \T bins.

In the bottom row of Fig. \ref{fig:martinez} we show the difference of the fractions in the upper row compared to the fractions of our field sample. At all \T, the QF is consistently above the EF at all stellar masses. As \T increases, the QF continues to deviate from the field, while it takes longer for the EF to deviate significantly. Specifically, at \T=2.77 Gyr the QF is already above the dashed line, while the EF takes 4.5 Gyr to differ from the field at all stellar masses. These results strengthen our previous conclusion that it takes a shorter time to quench a cluster galaxy than it does to transform its morphology from disc to bulge-dominated. These findings are consistent with \citet{Martinez2023}. 

\subsection{Bulge growth and disc fading quenching mechanisms}
\label{sec:bulgeMF}
\begin{figure*}
    \includegraphics[width=\textwidth]{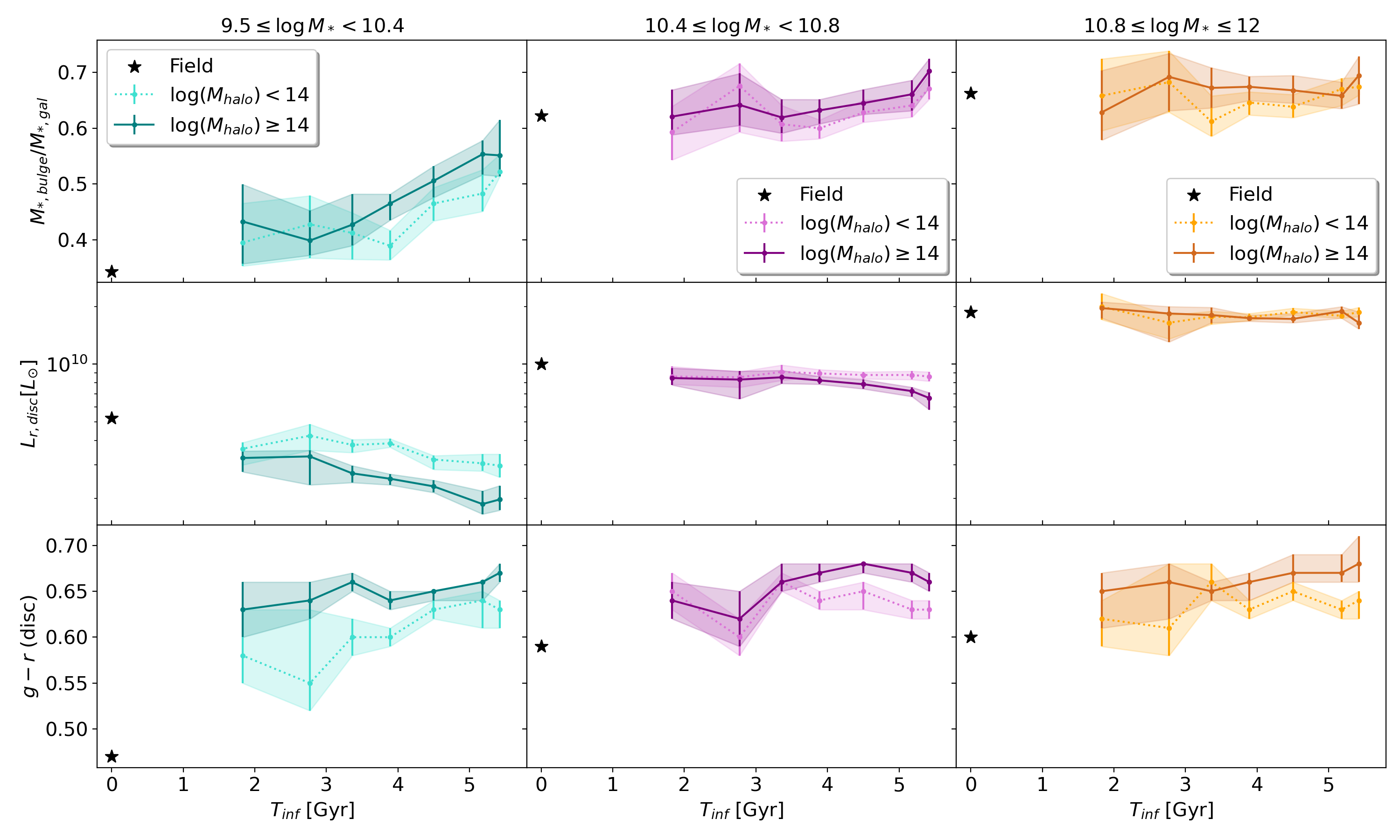}
    \caption{Top: The median bulge mass fraction of galaxies as a function of time since infall. Middle: The median r-band disc luminosity as a function of time since infall. Bottom: The median g-r disc colour as a function of time since infall. The separate panels and colours are the same as those defined in Fig. \ref{fig:QF}, with error bars showing the 90\% confidence interval from 1000 bootstrap iterations. Each data point is also \V weighted. The measurement uncertainties on the bulge mass, total mass, luminosity and colours are not taken into account in this figure.}
    \label{fig:bulgeMF}
\end{figure*}

It has been a topic of recent works whether it is bulge growth or disc fading that is responsible for the transformation of galaxy morphology in dense environments \citep[e.g][]{Hudson2010, Head2014, Quilley2022}. However, looking at B/T alone it is impossible to disentangle the effects of bulge growth from disc fading as both lead to an increase in B/T. In this section, we explore bulge mass fraction, disc luminosity, and disc colour as functions of time since infall to study how the bulge and disc behave separately, and to better constrain the mechanisms driving morphological change.

Bulge mass fraction ($M_{\text{bulge}}/M_\star$) provides a continuous number between 0 and 1 that specifies how bulge dominated the stellar mass of a galaxy is. For our work, we adopt the bulge and disc mass estimates from \citet{Mendel2014} who extended the catalogue of \citet{Simard2011} to include u-, i-, and j-band photometry. \citet{Mendel2014} determine bulge, disc, and total mass estimates from fitting bulge+disc and Sersic profile photometric decompositions. They note the total masses derived from the bulge+disc fit ($M_{B+D}$) can deviate from the sum of the independently derived bulge and disc masses ($M_B+M_D$) when photometric uncertainties are high. The authors suggest removing galaxies with overestimated bulge masses, identified where the offset between $M_{B+D}$ and $M_B+M_D$ is greater than $1\sigma$. We remove these galaxies from our sample and use $M_B+M_D$ when calculating the bulge mass fraction. However, we note that keeping these galaxies does not change our result, nor does using $M_{B+D}$ as the total galaxy mass instead of $M_B+M_D$. Cross matching our sample to this catalogue results in a final sample of 19112 galaxies in groups and clusters, and 63145 field galaxies. 
 
Bulge mass fraction as a function of \T is plotted in the top row of Fig. \ref{fig:bulgeMF}. The overall trends are similar to Fig. \ref{fig:BT}, where low mass galaxies in clusters show the strongest trend with \T. As galaxy stellar mass increases, the offset between galaxy groups and clusters becomes less pronounced and the trends with \T flatten. The low mass field galaxies have smaller bulge mass fractions than group and cluster galaxies, but field galaxies become increasingly similar to the group and cluster galaxies with increasing stellar mass. We explore disc fading in the bottom two rows of Fig. \ref{fig:bulgeMF}. Here, we plot the r-band disc luminosity ($L_{\text{r,disc}}$) and the g-r disc colour as functions of \T, two properties that are common when studying the effects of disc fading \citep[e.g.][]{Hudson2010, Quilley2022}. There is evidence of disc fading in the low mass galaxies, with luminosities decreasing and colours reddening with increasing time since infall. The trends of disc luminosity and colour with \T flatten at higher stellar masses (high mass field galaxies have more luminous and redder discs compared to ones of lower mass).

By looking at bulge mass fraction, disc luminosity and disc colour in Fig. \ref{fig:bulgeMF}, it appears both bulge growth and disc fading drive morphological transformation to some extent. Low mass galaxies in particular experience the strongest change over time in both bulge mass fraction and disc luminosity and colour. However, bulge growth appears to show the strongest trend for low mass galaxies compared to the disc luminosity and colour, suggesting the morphological transformation of low mass galaxies is primarily driven by bulge growth. The intermediate and high mass galaxies both show very little change in bulge mass fraction, disc luminosity, and disc colour over time so it is difficult to disentangle whether it is bulge growth or disc fading that is primarily responsible for changes in the morphology of these galaxies.

Several different environmental quenching mechanisms may be responsible for the cessation of star formation, the growth of galaxy bulges, and the fading of galaxy discs. Bulge-enhancing mechanisms tend to drive gas into the galaxy centre, possibly feeding a supermassive black hole and quenching star formation through AGN feedback in high mass galaxies \citep{Schawinski2006, Schawinski2007, Head2014}. Often, mergers within groups/in the field and repeating galaxy interactions within clusters can explain this scenario \citep{Christlein2004, Boquien2019, Quilley2022, Moore1996}, as well as the morphological disruption of galaxy discs. 

Mechanisms effecting the amount of gas in a galaxy's disc can lead to disc fading, and recently both RPS and starvation have been favoured as primary satellite quenching mechanisms \citep{McGee2009, Wetzel2015rps, Jeon2022}. RPS can directly strip cold gas from the disc, removing gas that could have otherwise formed stars. This process can quench galaxies on short timescales (< 1 Gyr, \citet{Quilis2000}) while having little to no effect on overall galaxy morphology. However, RPS is thought to play an important role in the fading of galaxy discs as it preferentially strips gas near the edges of a galaxy. This can lead to long HI tails and the apparent truncation of HI and H$\alpha$ discs \citep{Chung2007, Bekki2014}. Starvation results in a reduced accretion rate of halo gas onto the disc, limiting the amount of gas available for future star formation. This process is thought to be one of the dominant environmental process which begins in low mass halos, and quenches star formation on timescales $>2$Gyr \citep{McGee2009}. 

While a combination of bulge growth and disc fading mechanisms may drive morphological change for intermediate and high mass galaxies, the morphological evolution of low galaxies is primarily driven by bulge growth likely due to mergers and interactions.

\subsection{Pre-processing of star formation and morphology}
\label{sec:prepro}
In Section \ref{EFQF} we found that morphological transformation timescales are longer than quenching timescales, which are both shorter in clusters compared to groups. A common explanation for why quenching timescales are shorter in high density environments is pre-processing \citep[e.g.][]{delucia2012, Wetzel2013}. In this section, we investigate the effect of pre-processing on star formation and morphology for galaxies falling into groups and clusters.

\begin{figure}
    \centering
    \includegraphics[width=0.45\textwidth]{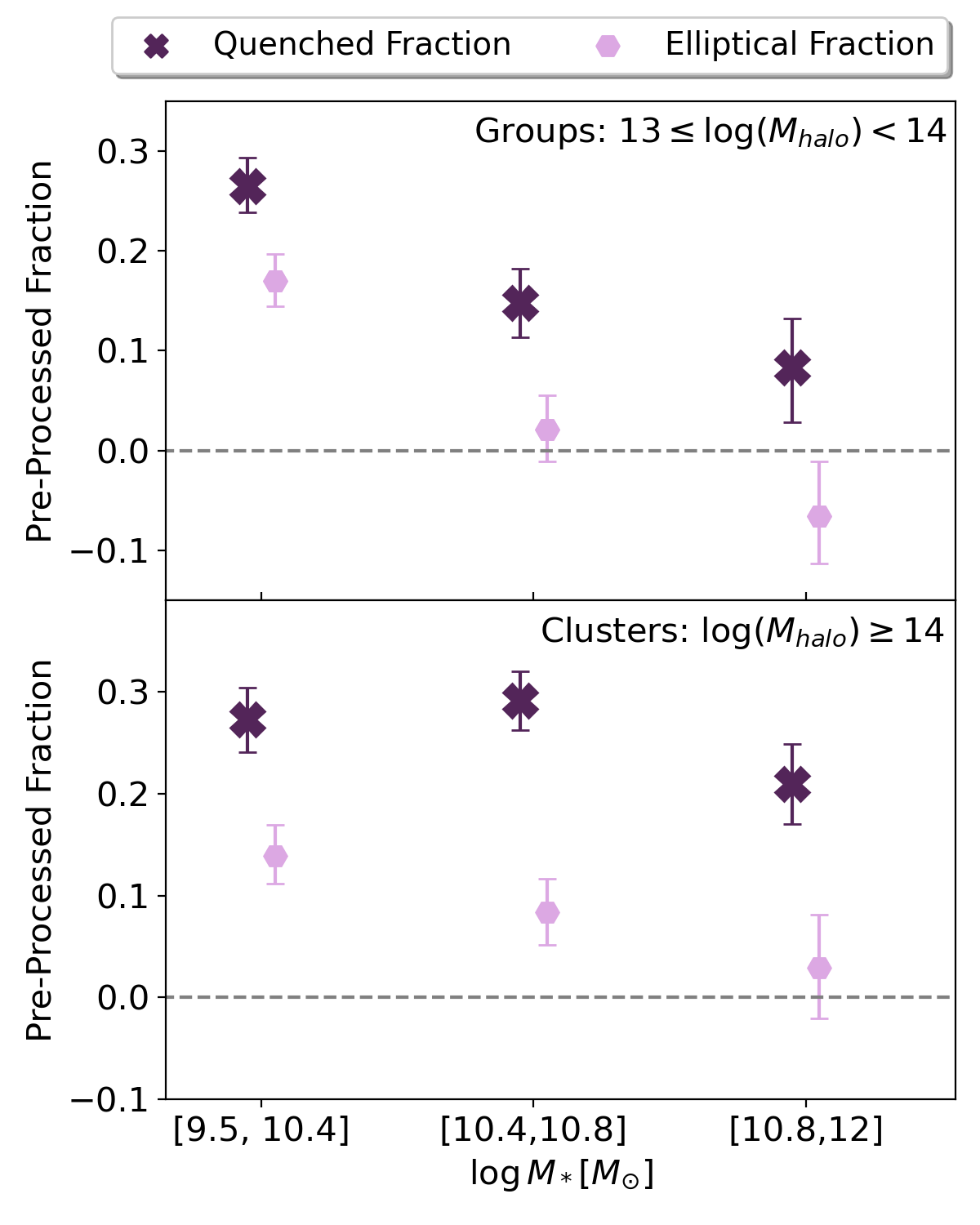}
    \caption{The pre-processed fraction for low, intermediate, and high mass galaxies in groups (top) and clusters (bottom). The pre-processed fraction associated with star formation and morphology are plotted as dark purple crosses and light purple hexagons, respectively. The errors represent the $1\sigma$ uncertainty.}
    \label{fig:pre-pro}
\end{figure}

To quantify the amount of pre-processing affecting SFR in this work, we calculate the difference in the field QF and the first \T bin from Fig. \ref{fig:QF} for each stellar mass and environment. We calculate the amount of pre-processing affecting morphology similarly, using the field EF and first \T data point in Fig. \ref{fig:EF}. The pre-processed fractions for group and cluster environments are shown in Fig. \ref{fig:pre-pro}. Overall, pre-processing has an effect on both morphology and star formation, yet star formation is more strongly pre-processed. We find the fraction of pre-processed galaxies in groups ranges from $\sim 10-30\%$ when considering the quenched fraction, and $\sim 0-20\%$ when considering the elliptical fraction. A negative pre-processed fraction, as is the case for high mass galaxies in groups, suggests these galaxies do not experience significant morphological pre-processing. The pre-processed fractions also decrease with increasing stellar mass, and both fractions are comparable or slightly larger for clusters.

Prior studies have attempted to constrain the effects of pre-processing either by tracing galaxy evolution through simulations or observationally by measuring substructure in galaxy clusters. \citet{Bahe2013} use the GIMIC suite of high-resolution hydrodynamic simulations and find the fraction of galaxies that have been satellites of a previous halo of mass $> 10^{13} M_{\odot}$ ranges from $<10\%$ to $60\%$ for galaxies currently in haloes of mass $\leq 10^{13.5} M_{\odot}$ and $10^{15.2} M_{\odot}$, respectively. Similarly, \citet{Han2018} use the YZiCS and find that $\sim 48 \%$ of today's cluster galaxies were once satellites of other hosts and $\sim 75 \%$ of these have experienced significant mass loss via tidal stripping due to pre-processing. Also using the YZiCS are \citet{Jung2018} who study the gas content of satellite galaxies in clusters with masses ranging from $5\times 10^{13} < M_{200}/M_{\odot} < 15$. They find evidence that ~34$\%$ of their galaxies are gas poor prior to entering their current cluster environment. \citet{Haines2015} model satellite infall applied to observations in an attempt to recreate the known trend of star-forming fraction and cluster-centric radius. The authors find that recently infalling galaxies have a star-forming fraction that is reduced by 19$\%$ compared to the field.

In terms of galaxy morphology, \citet{Wilman2009} study the fraction of S0-type galaxies in group and field environments at z$\sim$0.4. They find an excess of S0s at fixed luminosity in groups compared to the field, and beyond 0.3$h^{-1}_{75}$ Mpc from group centres. They find this hints at pre-processing playing an important role in the formation of S0 galaxies. More recently, \citet{Brambila2023} find a statistically significant difference in the asymmetry and smoothness parameters between field and cluster galaxies, finding evidence for pre-processing affecting galaxy morphology as well as star formation. 

\citet{Roberts2017} used observational data from the SDSS to make direct comparisons between the pre-processing of star formation and morphology. The authors quantify the degree to which star-forming fractions and disc fractions are suppressed in dense environments compared to the field for low and high mass galaxies. Qualitatively, they found evidence of both the pre-processing of star formation and morphology, although could not say for certain if one is more strongly pre-processed than the other. Quantitatively, they find the fraction of pre-processed low-mass galaxies ranges between 4-11$\%$ for star-forming fraction and 4-7$\%$ for disc fractions, which both decrease for higher mass galaxies. The largest pre-processed fractions are found in low mass galaxies falling into high mass haloes. Our results are generally in agreement qualitatively, although our low mass galaxies in clusters have similar pre-processed fractions compared to intermediate mass galaxies in clusters. This is not surprising looking at Fig. \ref{fig:QF}, and may be driven by large uncertainties in infall time. 

\section{Summary \& Conclusions}
\label{summary/conclusions}
In this work we investigate the environmental effect of galaxy groups and clusters on the star formation rate and morphology of 20369 galaxies in the Sloan Digital Sky Survey Data Release 7. We use location in projected phase space as a proxy for time since infall to study the timescales associated with star formation quenching and morphological transformation. We separate our galaxies into three stellar mass bins (low: $9.5 \leq\text{log}(M_{\star}/M_{\sun}) < 10.4$, intermediate: $10.4 \leq\text{log}(M_{\star}/M_{\sun}) < 10.8$, and high: $10.8 \leq\text{log}(M_{\star}/M_{\sun}) < 12$), and distinguish between galaxies in groups ($13 \leq \text{log}(M_{\text{halo}}/M_{\sun}) < 14$) and clusters ($\text{log}(M_{\text{halo}}/M_{\sun}) \geq 14$). We compare our results to a sample of 73474 field galaxies, and study how the quenched and elliptical fractions of galaxies change over time as galaxies fall into groups and clusters. Our main findings are as follows:
\begin{itemize}
    \item Changes in both the quenched fraction and elliptical fraction depend on stellar mass and environment, where we find clear evidence of increasing QF and EF as functions of \T. The strongest trends are found for low mass galaxies falling into clusters.
    \item By comparing the quenching and morphological transformation timescales in Section \ref{EFQF}, we find that morphological timescales are longer than quenching timescales in both groups and clusters. This supports the idea that quenching occurs faster than morphological transformation in both environments. However, low mass galaxies show no signs of significant quenching or morphological transformation in group environments (timescales are $\geq$ 9 Gyr). 
    \item By comparing the QF and EF of recently infalling galaxies to our field population in Section \ref{sec:prepro}, we find evidence for the pre-processing of both star formation rate and morphology, with pre-processing having a stronger effect on star formation. The pre-processing fractions typically decrease with increasing stellar mass, and slightly increase in clusters compared to groups
    \item Our analysis favours quenching mechanisms that act quickly to suppress sSFR, and act on longer timescales to grow galaxy bulges and fade galaxy discs. Possible mechanisms that quench star formation are RPS and starvation,  while mergers and galaxy interactions may drive morphological transformation.
\end{itemize}


\section*{Acknowledgements}
We would like to thank the anonymous referee for helpful comments and suggestions that improved this paper. LCP would like to thank the Natural Science and Engineering Council of Canada for funding. MO thanks M. Bravo, L. Foster, and D. Lazarus for helpful discussions. This work was made possible thanks in large part to number of publicly available software packages, including Astropy \citep{Astropy2022}, Matplotlib \citep{Hunter2007}, Numpy \citep{Harris2020}, Pandas \citep{McKinney2010}, SciPy \citep{Virtanen2020}, and Topcat \citep{Taylor2005}.

Funding for the Sloan Digital Sky Survey V has been provided by the Alfred P. Sloan Foundation, the Heising-Simons Foundation, the National Science Foundation, and the Participating Institutions. SDSS acknowledges support and resources from the Center for High-Performance Computing at the University of Utah. The SDSS web site is www.sdss5.org.

SDSS is managed by the Astrophysical Research Consortium for the Participating Institutions of the SDSS Collaboration, including the Carnegie Institution for Science, Chilean National Time Allocation Committee (CNTAC) ratified researchers, the Gotham Participation Group, Harvard University, The Johns Hopkins University, L’Ecole polytechnique fédérale de Lausanne (EPFL), Leibniz-Institut für Astrophysik Potsdam (AIP), Max-Planck-Institut für Astronomie (MPIA Heidelberg), Max-Planck-Institut für Extraterrestrische Physik (MPE), Nanjing University, National Astronomical Observatories of China (NAOC), New Mexico State University, The Ohio State University, Pennsylvania State University, Smithsonian Astrophysical Observatory, Space Telescope Science Institute (STScI), the Stellar Astrophysics Participation Group, Universidad Nacional Autónoma de México, University of Arizona, University of Colorado Boulder, University of Illinois at Urbana-Champaign, University of Toronto, University of Utah, University of Virginia, and Yale University.


\section*{Data Availability}
Galaxies used in this analysis were selected from the SDSS-DR7. Group and cluster membership was taken from the \href{https://gax.sjtu.edu.cn/data/Group.html}{Yang Group Catalogue}. Galaxy specific star formation rates were calculated from the star formation rates and stellar mass estimates of \href{https://salims.pages.iu.edu/gswlc/}{GSWLC-2}, while B/T morphology measurements came from the \href{http://alan-meert-website-aws.s3-website-us-east-1.amazonaws.com/fit_catalog/index.html}{SDSS PhotDec Catalogue}. Smoothness parameters are from \href{https://cdsarc.cds.unistra.fr/viz-bin/cat/J/ApJS/196/11}{Simard et. al. 2011}, and the bulge mass fractions were calculated from \href{https://cdsarc.cds.unistra.fr/viz-bin/cat/J/ApJS/210/3}{Mendel et. al. 2014}.


\bibliographystyle{mnras}
\bibliography{refs}



\appendix

\section{Star Formation Quenching}
\label{sSFR}

\begin{figure*}
    \includegraphics[width=\textwidth]{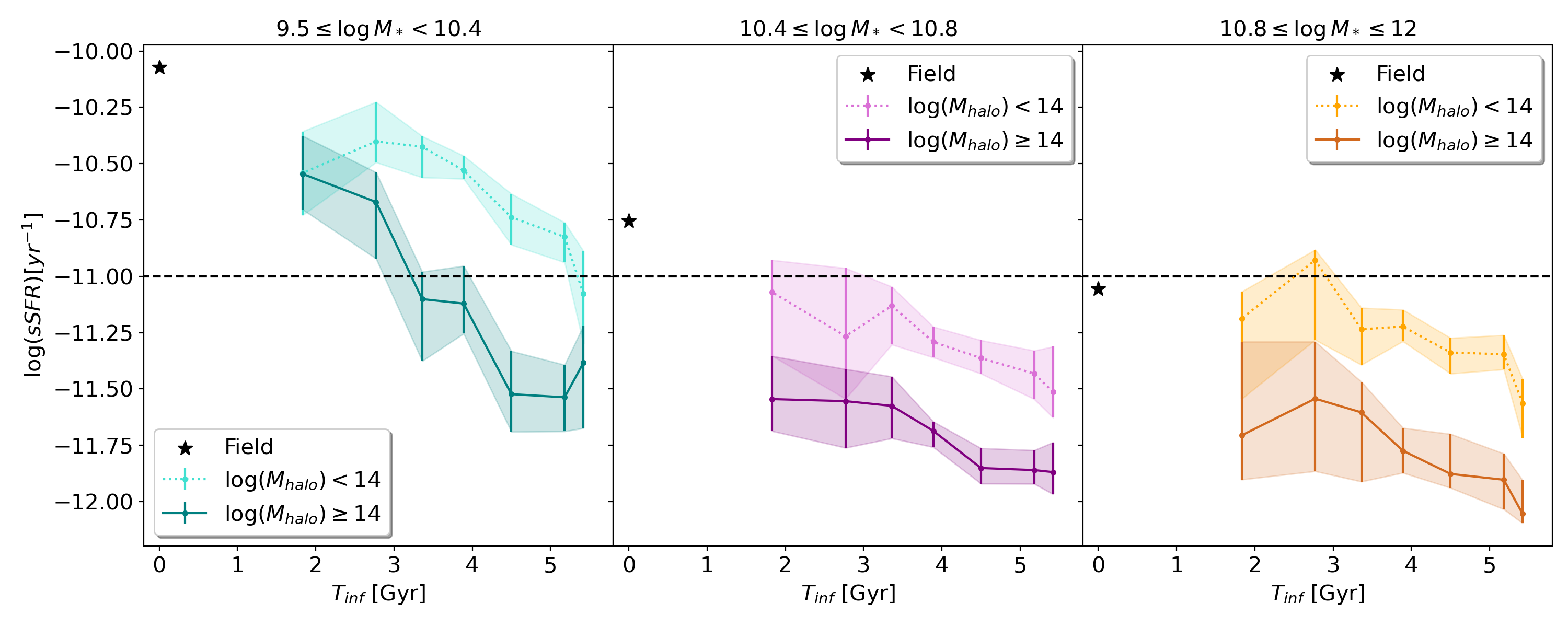}
    \caption{The 1/\V weighted median sSFR as a function of time since infall. The separate panels and colours are the same as those defined in Fig. \ref{fig:QF}. The error bars show the 90\% confidence interval from 1000 bootstrap iterations, and black dashed line at $\log(\text{sSFR})=-11 \text{yr}^{-1}$ represents our distinction between star-forming and quenched galaxies. The measurement uncertainty on sSFR is not taken into account in this figure.}
    \label{fig:sSFR}
\end{figure*}

To complement the quenched fraction shown in Fig. \ref{fig:QF}, in Fig. \ref{fig:sSFR} we plot median galaxy sSFR as a function of \T. Over time, sSFRs decrease as satellite galaxies fall into both groups and cluster environments. Trends are strongest for low mass galaxies falling into clusters.

\section{Morphological Transformation}
\label{BT}

\begin{figure*}
    \includegraphics[width=\textwidth]{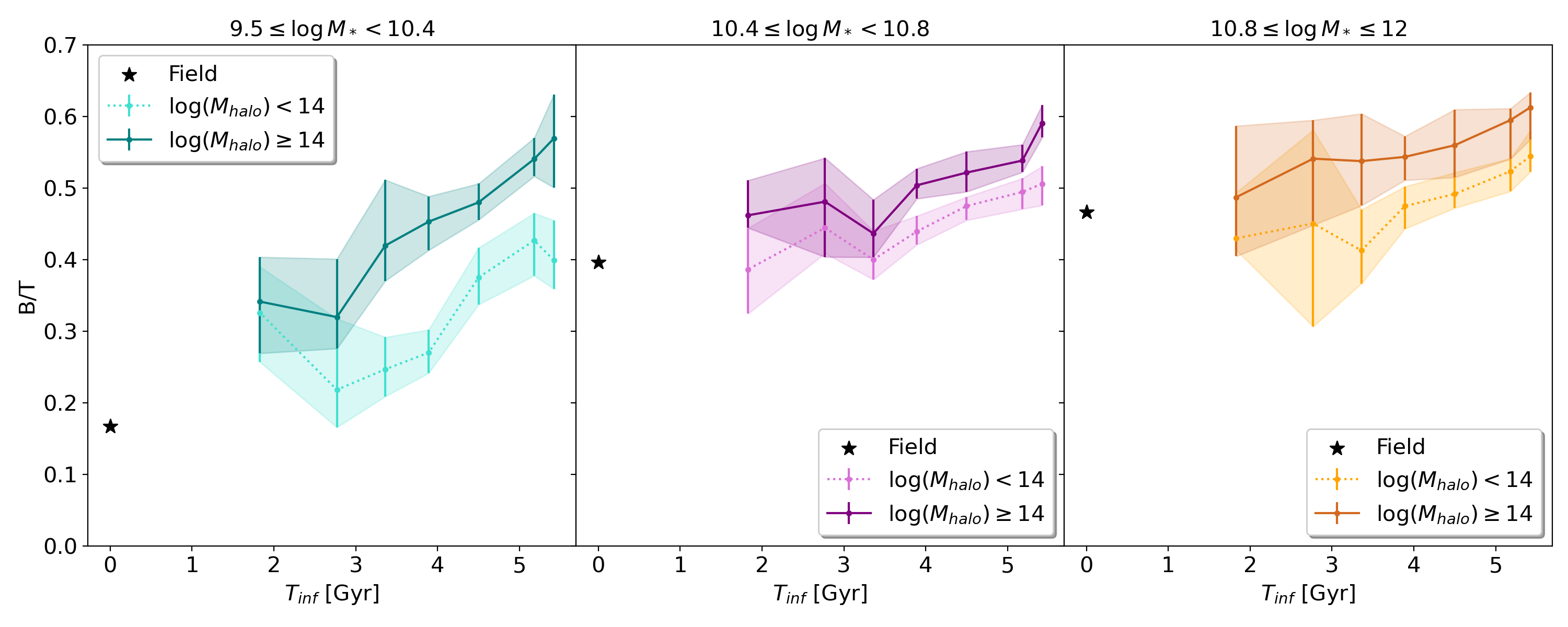}
    \caption{The 1/\V weighted median B/T as a function of time since infall. The separate panels and colours are the same as those defined in Fig. \ref{fig:QF}, with errors showing the 90\% confidence interval from 1000 bootstrap iterations. The measurement uncertainty on the B/T is not taken into account in this figure.}
    \label{fig:BT}
\end{figure*}

To complement the elliptical fraction shown in Fig. \ref{fig:EF}, in Fig. \ref{fig:BT} we plot the median B/T as a function of \T. As galaxies fall into groups and clusters, B/Ts increase and galaxies become more bulge dominated. Once again, the trends are strongest for low mass galaxies in clusters.

\section{B/T vs. Specific Star Formation Rate}
\label{BTsSFR}

\begin{figure*}
    \includegraphics[width=\textwidth]{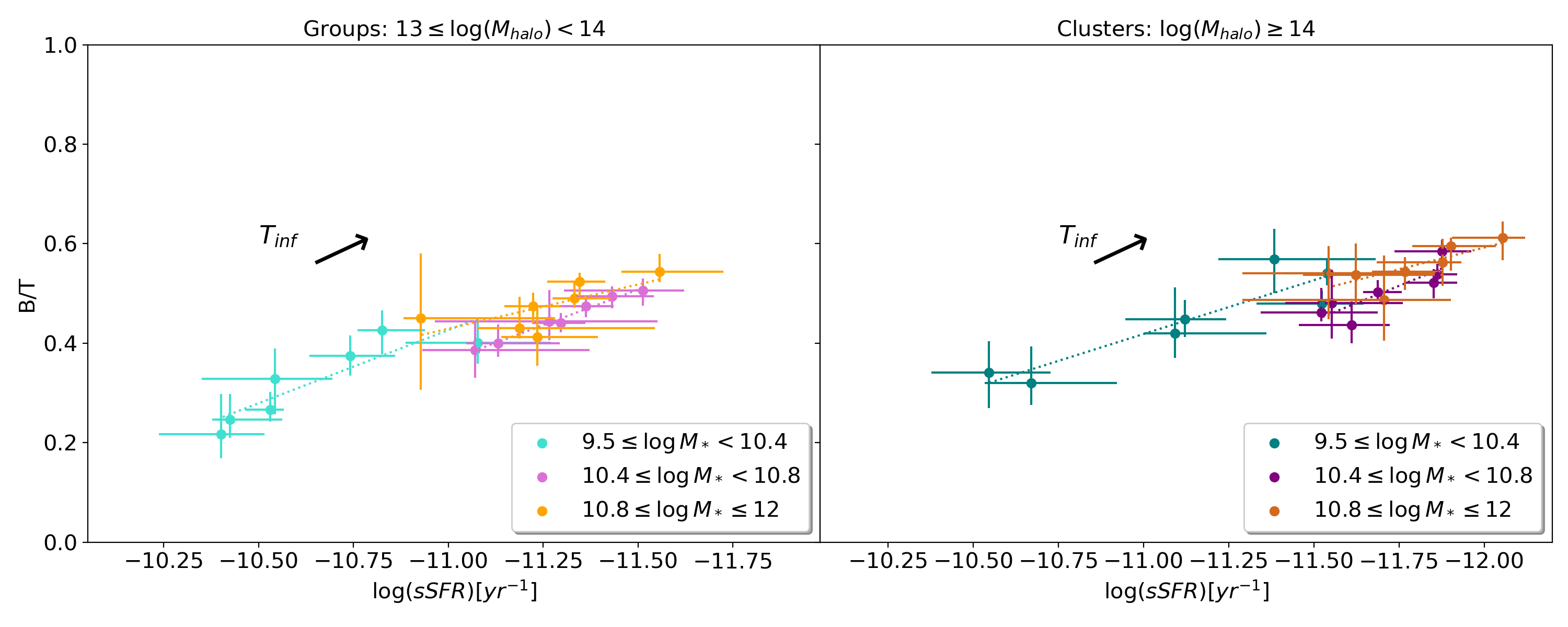}
    \caption{The 1/\V weighted median B/T versus sSFR. The left and right panels correspond to galaxies in groups and clusters, respectively, while galaxies in different stellar mass bins are represented by different coloured points shown in the legend. Time since infall increases from the bottom left to top right, represented by the illustrative black arrow. The linear least squares fit to each population is shown by dotted lines, and the error bars represent the 90\% confidence intervals from 1000 bootstrap iterations.}
    \label{fig:BTsSFR}
\end{figure*}

In Fig. \ref{fig:BTsSFR}, we show B/T versus sSFR for galaxies in groups (left) and clusters (right). As stated in Section \ref{EFQF}, a direct comparison between B/T and sSFR can be hard to interpret given the different units, but plotted in this way enables us to make comparisons to fig. 10 of \citet{Sampaio2022}. Over time, B/T increases while sSFR decreases, as expected. It appears there is a greater change in sSFR compared to B/T, suggesting quenching occurs prior to significant changes in morphology.

\section{Timescale Stability Test}
\label{stability test}
As stated in Section \ref{sec:prior}, the results of our analysis are dependent on the specific thresholds chosen to define elliptical and quenched galaxies. In this appendix, we preform a stability test to explore how our results (that SFR change more quickly than morphology) change if we chose different (but reasonable) B/T and sSFR thresholds.

\begin{figure*}
    \includegraphics[width=0.45\textwidth]{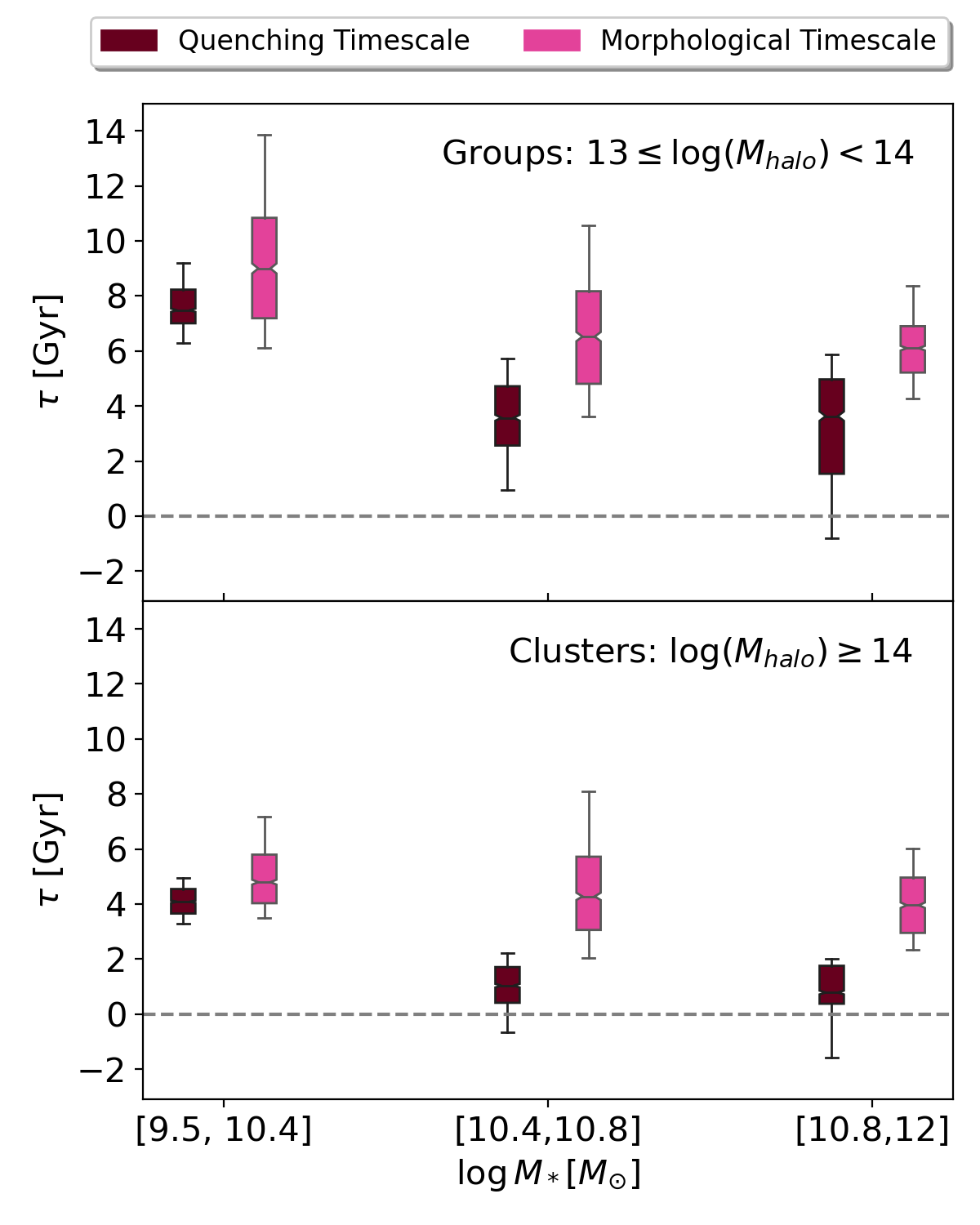}
    \caption{The distribution of quenching and morphological transformation timescales resulting from the selection of different thresholds chosen to define elliptical and quenched galaxies. Timescales are plotted for low ($9.5 \leq\text{log}(M_{\star}/M_{\sun}) < 10.4$), intermediate ($10.4 \leq\text{log}(M_{\star}/M_{\sun}) < 10.8$), and high ($10.8 \leq\text{log}(M_{\star}/M_{\sun}) < 12$) mass galaxies in groups (top) and clusters (bottom). The quenching timescales are in burgundy, while the morphological transformation timescales are plotted in pink. The inner filled region represents the 20-75 percentiles and the whiskers show the full extent of the data.}
    \label{fig:stability}
\end{figure*}

We perform this test by randomly selecting new thresholds of B/T and sSFR between 0.4<B/T<0.6 and -11<log(sSFR)<-11.5 for our definition of elliptical and quenched, respectively. This is performed 1000 times, and each time new quenching and morphological transformation timescales are calculated for each galaxy population following the same procedure as in Section \ref{EFQF}. The distribution of timescales is shown in Figure \ref{fig:stability}.

Quenching and morphological timescales are clearly dependent on the thresholds chosen when defining quenched and elliptical galaxies. The larger the B/T threshold chosen, the longer the morphological timescale. The larger the sSFR chosen, the lower the quenching timescale is. However, the morphological timescales are still above the quenching timescales in general, supporting our original conclusion that quenching happens prior to morphological transformation.


\bsp	
\label{lastpage}
\end{document}